# Evidence Contrary to the Existing Exo-Planet Migration Concept

J. Marvin Herndon
Transdyne Corporation
San Diego, CA 92131 USA

December 28, 2006

**Communications:** mherndon@san.rr.com    http://UnderstandEarth.com

## Abstract

Exo-planet migration is assumed to have occurred to explain close-to-star gas giant exo-planets within the context of the so-called standard model of solar system formation, rather than giving cause to question the validity of that particular model. I present evidence against the concept of planet migration, evidence that is historical, interdisciplinary, and model-independent. First, I demonstrate a flaw in the standard model of solar system formation that would lead to the contradiction of terrestrial planets having insufficiently massive cores. Then, I discuss the evidence that points to the Earth previously having been a Jupiter-like close-to-Sun gas giant and the consequences that arise there from. Observations of close-to-star gas giant exo-planets orbiting stars other than our own Sun, rather than being evidence for planet migration, I submit, are evidence for differing degrees of violence associated with the thermonuclear ignition of their particular stars. As observational resolution improves, one might expect to find other exo-planetary systems, like our own, with their innermost planets stripped of gaseous envelopes. Such systems may be a necessary requirement for the existence of life.

## Introduction

Recent discoveries of close-to-star gas giant exo-planets in planetary systems other than our own Solar System has led to an interesting astrophysical debate. Generally, in attempting to explain those observations, the astronomical community is presented with choices, among those being the following: (1) Attempt to explain new exo-planet observations within the context of currently existing models developed to explain our own Solar System; or (2) Utilize exo-planet observations as a reference to question what might be wrong with extant Solar System models. The widely adopted assumption of exo-planet migration as an explanation of close-to-star gas giants is an excellent example of (1) above to the exclusion of (2). Here, I present evidence against the concept of exo-planet migration, evidence that is historical, interdisciplinary, and non-model-dependant. First I demonstrate a flaw in the so-called standard model of solar system formation. Then I discuss the evidence that points to the Earth previously having been a Jupiter-like close-to-Sun gas giant non-exo-planet.

There have long been mainly two ideas about how the planets of the Solar System formed. In the 1940s and 1950s, the idea was discussed about planets "raining out" from inside of giant gaseous proto-planets with hydrogen gas pressures on the order of $10^2$-$10^3$ bar (Eucken 1944; Kuiper 1951; Urey 1951). But, in the early 1960s, astrophysicists specifically began making models of primordial matter, not forming dense protoplanets, but rather spread out into a very low-density "solar nebula" with hydrogen gas pressures on the order of $10^{-5}$ bar. The idea of low-density planetary formation, often referred to as the standard model, envisioned that dust would condense at fairly low temperatures, and then would gather into progressively larger grains, and become rocks, then planetesimals, and ultimately planets (Stevenson 1982; Wetherill 1980).

## Standard Model of Solar System Formation in Error

There are three principal types of chondritic meteorites: Carbonaceous, Enstatite, and Ordinary. I have shown by ratios of mass that the whole Earth (and especially the endoEarth, the inner 82% of its planetary bulk) is like an enstatite chondrite.

Imagine melting a chondrite in a gravitational field. At elevated temperatures, the iron metal and iron sulfide components will alloy together, forming a dense liquid that will settle beneath the silicates like steel on a steel-hearth. The Earth is like a spherical steel-hearth with a fluid iron-alloy core surrounded by a silicate mantle.

The Earth's core comprises about 32.5% by mass of the Earth as a whole. Only the enstatite chondrites, not the ordinary chondrites, have the sufficiently high proportion of iron-alloy that is observed for the core of the Earth, as shown in Fig. 1.

The two main hypotheses about planetary formation, at very high pressures and at very low pressures, embody fundamentally different condensation processes which appear to be the underlying cause for the strikingly different states of oxidation of enstatite and carbonaceous chondrite matter. Matter like that of enstatite chondrites has the state of oxidation expected to have resulted from raining out at high temperatures and high pressures (Eucken 1944; Herndon 2004c, 2006b; Herndon & Suess 1976), whereas ordinary chondrites appear to have formed from a mix of those two types of matter (Herndon 2004a, 2006b).

The extant standard model of solar system formation envisions condensation occurring at very low pressures, hence at low temperatures. Consequently, the state of oxidation of would be like that of carbonaceous chondrites, having little or no metallic iron. Instead of the metal, iron would occur mainly as magnetite, $Fe_3O_4$. The standard model of solar system formation is wrong because it would lead to terrestrial planets having insufficiently massive cores.

One is thus left with the idea of planets raining out from within a giant gaseous protoplanet (Eucken 1944), which appears to explain the high state of reduction observed for enstatite chondrites and for the Earth (Herndon 2004b, 2004c; Herndon & Suess 1976).

## Earth as a Close-to-Star Jupiter-like Gas Giant

Planets generally consist of more-or-less uniform, closed, concentric shells of matter, layered according to density. The crust of the Earth, however, is an exception. Approximately 29% of the surface area of the Earth is composed of the portions of continents that presently lie above mean sea level; an additional 12% of the surface area of the Earth is composed of the continental margins, which are submerged to depths of no more than 2 km (Mc Lennan 1991). The continental crust is less dense and different in composition than the remaining surface area, which is composed of ocean-floor basalt.

To date there has been no adequate geophysical explanation to account for the formation of the non-contiguous, crustal continental rock layer, except the idea put forth in 1933 by Hilgenberg (1933) that in the distant past for an unknown reason or reasons the Earth had a smaller diameter and, consequently, had a smaller surface area. From modern surface area measurements, the smaller radius required would be about 64% of its current radius, which would yield a mean density for the Earth of 21 g/cm$^3$. The reason for Earth's smaller radius, I submit, is that the Earth rained out from within a giant gaseous protoplanet and originally formed as the rock-plus-alloy kernel of a close-to-star giant gaseous planet like Jupiter (Herndon 2004b, 2004c, 2006b).

The mass of protoplanetary-Earth, calculated from solar abundance data (Anders & Grevesse 1989) lies in the range of about 275 to 305 times the mass of the present-day Earth. That mass is quite similar to Jupiter's mass, 318$m_E$.

Pressures at the gas-rock boundary within the interior of Jupiter are estimated to be in the range from 43 Mbar to 60 Mbar (Podolak & Cameron 1974; Stevenson & Salpeter 1976). Such pressure in a terrestrial, Jovian-like gas envelope is sufficient to compress the protoplanetary alloy-plus-rock core that became the Earth to a mean density of about 21 g/cm$^3$, a value virtually identical to that expected for a smaller Earth with a contiguous, closed, crustal continental shell. That identity, I submit, stands as evidence of the Earth having been a giant, gaseous planet like Jupiter (Herndon 2004b, 2004c, 2006b).

## Consequences and Implications

Hilgenberg's (1933) idea that the Earth initially was smaller in diameter, without oceans, and with continental-rock forming a uniform, closed shell of matter covering the entire surface of the planet, which subsequently fragmented into continents upon expansion, separated by ocean basins of new oceanic crust, became the basis for Earth expansion theory (Carey 1976, 1988; Hilgenberg 1933; Scalera 1990; Scalera & Jacob 2003). Lack of knowledge of a means for Earth compression and/or of an energy source for Earth expansion was a major impediment for that geodynamic theory (Beck 1961; Cook & Eardley 1961; Jordan 1971). The alternative view, plate tectonics theory, developed in the 1960s and which seems to explain well topographic features of the ocean floor, suffers from the impediment of necessitating solid-state mantle convection for which no unambiguous evidence exists and does not provide an energy source for geodynamics. Recently, I published the unification of these two seemingly disparate theories, that I call *Whole-Earth Decompression Dynamics*, which obviates the extant endemic problems (Herndon 2005b).

The principal consequences of Earth's formation from within a giant gaseous protoplanet are profound and affect virtually all facets of Geophysics in fundamental ways (Herndon 2006b). Principal implications result (*i*) from Earth having been compressed by about 300 Earth-masses of primordial gases which provides a major source of energy for geodynamic processes, and (*ii*) from the deep-interior having a highly-reduced state of oxidation which results in great quantities of uranium and thorium existing within the Earth's core (Herndon 2005a), and leads to the feasibility of the georeactor, a hypothesized natural, nuclear fission reactor at the center of the Earth as the energy source for the geomagnetic field (Herndon 1993, 1994, 1996, 2003).

After being stripped of its great, Jupiter-like atmospheric overburden of volatile protoplanetary constituents, presumably by the high temperatures and/or by the violent activity, such as T Tauri-phase solar wind (Joy 1945; Lada 1985; Lehmann, Reipurth, & Brander 1995), associated with the thermonuclear ignition of the Sun, the Earth would inevitably begin to decompress, to rebound toward a new hydrostatic equilibrium. The initial whole-Earth decompression is expected to result in a global system of major *primary* cracks appearing in the rigid crust which persist and are identified as the global, mid-oceanic ridge system, just as explained by Earth expansion theory. But here the similarity with that theory ends. Whole-Earth Decompression Dynamics sets forth a different mechanism for whole-Earth dynamics which involves the formation of *secondary* decompression cracks and the in-filling of those cracks yielding the oceanic features previously attributed to plate tectonics.

One of the consequences of Earth formation as a giant, gaseous, Jupiter-like planet, as described by Whole-Earth Decompression Dynamics, is the existence of a vast reservoir of energy, the stored energy of protoplanetary compression, available for driving geodynamic processes related to whole-Earth decompression. Some of that energy, I have suggested, is emplaced as heat at the mantle-crust-interface at the base of the crust through the process of *mantle decompression thermal-tsunami* (Herndon 2006a). Mantle decompression thermal-tsunami poses a basis for understanding the observed geothermal gradient in the crust and provides a new explanation for a portion of the internal heat being lost from the Earth. It may prove as well to be a significant energy source for earthquakes and volcanism, as these geodynamic processes appear concentrated along secondary decompression cracks.

Only three processes, operant during the formation of our Solar System, are responsible for the diversity of matter in the Solar System and are directly responsible for planetary internal-structures, including planetocentric nuclear fission reactors, and for dynamical processes, including and especially, geodynamics. These processes are: (*i*) Low-pressure, low-temperature condensation from solar matter in the remote reaches of the Solar System or in the interstellar medium; (*ii*) High-pressure, high-temperature condensation from solar matter associated with planetary-formation by raining out from the interiors of giant-gaseous protoplanets, and; (*iii*) Stripping of the primordial volatile components from the inner portion of the Solar System by super-intense solar wind associated with T-Tauri phase mass-ejections, presumably during the thermonuclear ignition of the Sun (Herndon 2006b).

Observations of close-to-star gas giant exo-planets orbiting stars other than our own, rather than being evidence for planet migration, I submit, are evidence for differing degrees of violence associated with the thermonuclear ignition of their stars (Herndon 1994). As observational resolution improves, one might expect to find other exo-planetary systems, like our own, with their innermost planets stripped of gaseous envelopes. Such systems may be a necessary requirement for the existence of life.

**Figure Caption**

**Fig. 1.** The percent mass of the alloy component of each of nine enstatite chondrites and 157 ordinary chondrites. This figure clearly shows that, if the Earth is chondritic in composition, the Earth as a whole, and especially the endo-Earth, is like an enstatite chondrite and *not* like an ordinary chondrite. The reason is clear from the abscissa which shows the molar ratio of oxygen to the three major elements with which it combines in enstatite chondrites and in ordinary chondrites. This figure adapted from Herndon (2006b) also clearly shows that, if the Earth is chondritic in composition, the Earth as a whole, and especially the endo-Earth, has a state of oxidation like an enstatite chondrite and *not* like an ordinary chondrite. Data from (Baedecker & Wasson 1975; Jarosewich 1990; Kallemeyn et al. 1989; Kallemeyn & Wasson 1981).

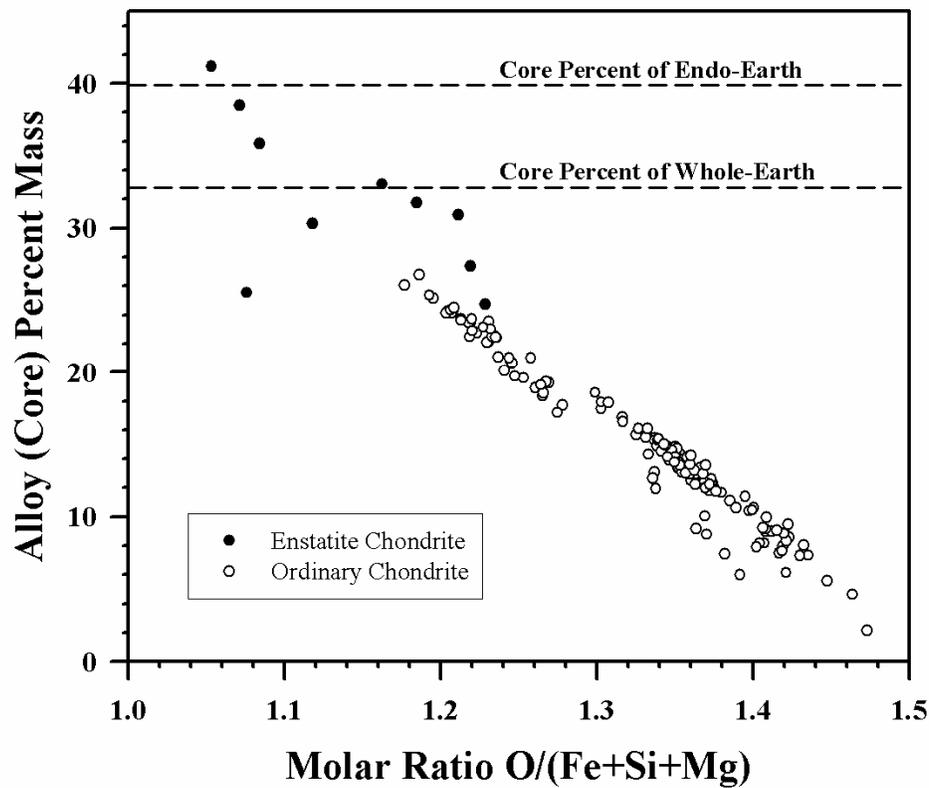